\documentclass[12pt]{article}
\usepackage{amsmath}
\usepackage{graphicx}
\usepackage{enumerate}
\usepackage{natbib}

\usepackage{amssymb, enumerate, amsthm}
\usepackage{bm}

\usepackage{booktabs}
\usepackage{siunitx}

\usepackage{algorithm} 
\usepackage{algpseudocode}

\usepackage{subcaption}

\newcommand{\blind}{1}

\newcommand{\Ib}{\bm{I}}
\newcommand{\Sb}{\bm{S}}
\newcommand{\Sigmab}{\bm{\Sigma}}
\newcommand{\ub}{\bm{u}}
\newcommand{\Ub}{\bm{U}}
\newcommand{\vb}{\bm{v}}
\newcommand{\Vb}{\bm{V}}
\newcommand{\xb}{\bm{x}}
\newcommand{\Xb}{\bm{X}}

\newcommand{\Epsilonb}{\bm{\mathrm{E}}}

\usepackage{xcolor}

\DeclareMathOperator*{\argmin}{arg\,min}

\DeclareMathOperator{\diag}{diag}
\DeclareMathOperator*{\rank}{rank}
\DeclareMathOperator*{\supp}{supp}

\theoremstyle{plain}

\newtheorem{Remark}{Remark}

\newtheorem{Corollary}{Corollary}

\begin{document}

\def\spacingset#1{\renewcommand{\baselinestretch}%
{#1}\small\normalsize} \spacingset{1}


\if1\blind
{
  \title{\bf High-Dimensional Block Diagonal Covariance Structure Detection Using Singular Vectors}
  \author{Jan O. Bauer\\
    Department of Econometrics and Data Science, Vrije Universiteit Amsterdam, \\
    Amsterdam, The Netherlands}
  \maketitle
} \fi

\if0\blind
{
  \bigskip
  \bigskip
  \bigskip
  \begin{center}
    {\LARGE\bf High-Dimensional Block Diagonal Covariance Structure Detection Using Singular Vectors}
\end{center}
  \medskip
} \fi

\bigskip
\begin{abstract}
The assumption of independent subvectors arises in many aspects of multivariate analysis. In most real-world applications, however, we lack prior knowledge about the number of subvectors and the specific variables within each subvector. Yet, testing all these combinations is not feasible. For example, for a data matrix containing 15 variables, there are already $1\,382\,958\,545$ possible combinations. Given that zero correlation is a necessary condition for independence, independent subvectors exhibit a block diagonal covariance matrix. This paper focuses on the detection of such block diagonal covariance structures in high-dimensional data and therefore also identifies uncorrelated subvectors. Our nonparametric approach exploits the fact that the structure of the covariance matrix is mirrored by the structure of its eigenvectors. However, the true block diagonal structure is masked by noise in the sample case. To address this problem, we propose to use sparse approximations of the sample eigenvectors to reveal the sparse structure of the population eigenvectors. Notably, the right singular vectors of a data matrix with an overall mean of zero are identical to the sample eigenvectors of its covariance matrix. Using sparse approximations of these singular vectors instead of the eigenvectors makes the estimation of the covariance matrix obsolete. We demonstrate the performance of our method through simulations and provide real data examples. Supplementary materials for this article are available online.
\end{abstract}

\noindent%
\vfill

\newpage
\spacingset{1.9} 
\section{Introduction}
\label{s:Introduction}

Let $\Sigmab$ be the population covariance matrix of a $p$-dimensional random vector. Estimating and testing the structure of $\Sigmab$ and $\Sigmab^{-1}$ is important in numerous real-world applications. However, in high-dimensional contexts where the number of variables exceeds the number of observations, estimating and testing become challenging. Such scenarios arise, for example, in the analysis of DNA microarray gene expressions or in the detection of block diagonal structures prior to network inference for Gaussian graphical models (see, e.g., \citet{FL06}, \citet{TWS15}, and related references). Consequently, there is a considerable interest in testing covariance matrices of high-dimensional data for a block diagonal structure. The objectives of these procedures can be categorized into two primary areas:

\begin{enumerate}
    \item Tests for mutually uncorrelated random variables, specifically, testing for a diagonal covariance structure $\Sigmab = \diag(\sigma_1^2,\ldots,\sigma_p^2)$ with unknown but finite positive constants $\sigma_1^2 ,\ldots, \sigma_p^2$ on the diagonal.

    \item Testing for uncorrelated subvectors, i.e., assessing if the covariance matrix exhibits a block diagonal structure $\Sigmab = \diag(\Sigmab_1, \ldots, \Sigmab_b)$.
\end{enumerate}

When testing the hypothesis $\Sigmab = \diag(\sigma_1^2,\ldots,\sigma_p^2)$, \citet{BJYZ09} and \citet{JY13} extended classical likelihood ratio tests to the high-dimensional context. Alternative methods are based on the distance between the sample covariance matrix and the diagonal matrix $ \diag(\sigma_1^2,\ldots,\sigma_p^2)$. Contributions in this category include \citet{LW02}, \citet{BD05}, \citet{FSG10}, and \citet{CZZ10}. Furthermore, \citet{JO01, JO08} used the distributional properties of the largest eigenvalue of the sample covariance matrix to construct hypothesis tests for sphericity. In addition, approaches for $\mathcal{L}_2$-type test statistics have been explored (see, e.g., \cite{SC05} and \citet{LD18}).

In the context of testing the hypothesis $\Sigmab = \diag(\Sigmab_1, \ldots, \Sigmab_b)$, \citet{JY13} also extended the likelihood ratio approach for this hypothesis to high-dimensional data, and \citet{JBZ13} proposed a corrected likelihood ratio test and a large-dimensional trace criterion. Among others, \citet{SR12}, \citet{HSNP15}, \citet{YA16}, and \citet{YHN17} used empirical distances between the covariance matrix and its block diagonal form to derive test statistics. In addition, \citet{BHPZ17} extended the Schott type test \citep{SC05} to test for independence of random vectors, and \citet{SR13} provided a test based on an extension of the distance correlation \citep{SRB07} to high-dimensional contexts.

In most applications, however, we lack prior knowledge about the number of blocks and the specific variables within each block. The number of possible distinct combinations of $p$ random variables into different blocks of unknown size can be calculated using the Bell number. For example, with $p=15$ variables, there are already around $1\,382\,958\,545$ possible combinations. Yet, testing all these combinations in high-dimensional contexts is not feasible due to the increasing risk of test errors.

Therefore, methods for detecting block diagonal structures of covariance matrices are needed. So far, \citet{PBT12}, \citet{TWS15}, and \citet{DG18} have proposed approaches to address this problem. \citet{PBT12} obtained block sparsity by extending the graphical lasso \citep{FHT07} which applies $\mathcal{L}_1$ regularization on the entries of the estimated covariance matrix. \citet{TWS15} recognized that the first step of the graphical lasso involves single linkage hierarchical clustering of the variables. However, single linkage clustering can yield suboptimal results in finite-sample contexts. To address this limitation, they introduced the cluster graphical lasso, which uses alternative linkages than single linkage for clustering. \citet{DG18} took a different approach. To detect block diagonal structures, they select the best fitting model from a collection of multivariate distributions with block diagonal covariance matrices. This collection of models is generated through hard-thresholding of the sample covariance matrix.

In this article, we contribute a novel nonparametric approach for detecting block diagonal structures. We exploit that the structure of the right singular vectors of the mean-centered data matrix mirror the block diagonal structure of the covariance matrix. This mirroring feature allows us to uncover the structure of the data matrix effectively. In practice, however, sample noise in the data matrix masks the block diagonal structure, consequently masking also the sample singular vectors. We propose employing sparse singular vectors, named \textit{sparse loadings}, to reveal the block diagonal structure. These sparse loadings are sparse approximations, i.e., have many zero values, of the sample singular vectors and can reflect the inherent structure of the population singular vectors. Their computation aligns with lasso-type regression, presenting a convenient pathway for employing the Bayesian information criterion to identify sparse loadings that represent the structure inherent in the population singular vectors.

The rest of this article is organized as follows: In Section~\ref{s:BDSVD}, after introducing basic notations and definitions, we present the concept for block diagonal covariance matrix detection using singular vectors. We provide elaborate simulation studies in Section~\ref{s:Simulations}, and real data examples in Section~\ref{s:Examples}. Section~\ref{s:Discussion} contains the discussion.

\section{Singular Vectors to Detect Block Diagonal Structure}\label{s:BDSVD}

\subsection{Preliminaries and Basic Ideas}

Let $\Xb \equiv (\xb_1,\ldots,\xb_p) \equiv  (\Xb_1,\ldots,\Xb_b)$ denote an $n \times p$ matrix of data. $\Xb$ contains $n$ observations, $p$ variables, $n \times 1$ column vectors $\xb_1,\ldots,\xb_p$, and is partitioned into $b$ distinct submatrices: $\Xb_i$ of dimension $n \times p_i$ for $i\in\{1,\ldots,b\}$ where $p = p_1 + \ldots + p_b$. Each $\Xb_i$ is organized such that $\Xb_1 = (\xb_1,\ldots, \xb_{p_1})$ contains the first $p_1$ columns of $\Xb$, $\Xb_2$ contains the next $p_2$ columns of $\Xb$, and so forth. We assume the convenient ordering $\Xb \equiv  (\Xb_1,\ldots,\Xb_b)^\top$ since this order can be obtained by using row permutation. The singular value decomposition (SVD) of the data matrix can be written as follows:
\begin{equation*}
    \Xb = \Ub \bm{D} \Vb^\top, \; \Ub^\top\Ub = \bm{I}_n, \; \Vb^\top \Vb = \bm{I}_p, \; d_1 \geq \ldots \geq d_r > 0 \, ,
\end{equation*}
where $r \leq \min(n,p)$ is the rank of $\Xb$, and $\Ub = (\ub_1,\ldots,\ub_n)$ and $\Vb = (\vb_1,\ldots,\vb_p)$ are the left and right singular vectors respectively. Without loss of generality, assume that the overall mean of $\Xb$ is zero. This implies that the right singular vectors of the data matrix are the eigenvectors of the sample covariance matrix.

We assume that the population covariance matrix $\Sigmab$ follows a block diagonal structure:
\begin{equation*}
 \Sigmab \equiv   \begin{pmatrix}
    \Sigmab_1 & \bm{0} & \bm{0} \\
    \bm{0} & \ddots & \bm{0}\\
    \bm{0} & \bm{0} & \Sigmab_b
    \end{pmatrix} \, ,
\end{equation*}
where $\Sigmab_i$ is the population covariance matrix corresponding to the submatrix $\Xb_i$ for $i\in\{1,\ldots,b\}$, and the population covariance between these submatrices is $\bm{0}$. In practice, we have only access to the sample covariance matrix of the observed data, which we denote by $\Sb$. The sample covariance matrix can be expressed by the population covariance matrix, perturbed by a noise matrix $\Epsilonb$,
\begin{equation*}
\Sb \equiv \Sigmab + \Epsilonb  \, ,
\end{equation*}
which masks the block diagonal structure of $\Sigmab$.

Assume for now that $\Epsilonb = \bm{0}$ such that $\Sb$ exhibits the block diagonal structure of $\Sigmab$. It holds that the right singular vectors of $\Xb$ mirror the block diagonal structure of the covariance matrix:
\begin{Corollary}\label{col:VectorStructure}
$\Sb$ is a block diagonal matrix iff its eigenvectors (the right singular vectors of $\Xb$) exhibit the structure
\begin{equation}\label{eq:VectorBlockStructure}
    \Vb \bm{P}_\pi =  \begin{pmatrix}
    \Vb_1 & \bm{0} & \bm{0} \\
    \bm{0} & \ddots & \bm{0}\\
    \bm{0} & \bm{0} & \Vb_b
    \end{pmatrix} \, ,
\end{equation}
where $\Vb_i$ are the eigenvectors of $\Sb_i$ (the right singular vectors of $\Xb_i$) for $i \in \{1,\ldots,b\}$, and $\bm{P}_\pi$ is a permutation matrix leading to a block diagonal structure.
\end{Corollary}

Consequently, by analyzing the structure of the singular vectors, it is possible to identify the block diagonal pattern of $\Sb$ and thus detect the presence of uncorrelated submatrices.

We saw that the singular vectors exhibit a structure that mirrors the block diagonal structure when $\Epsilonb = \bm{0}$. Under more realistic conditions when $\Epsilonb \neq \bm{0}$, however, the singular vectors are also perturbed by $\Epsilonb$. Let $\lambda_1 \geq \ldots \geq \lambda_p$ be the sample eigenvalues of $\Sb$, and $\tilde{\lambda}_1\geq \ldots \geq \tilde{\lambda}_p$ be the population eigenvalues with corresponding population eigenvectors $\tilde{\vb}_1,\ldots,\tilde{\vb}_p$ of $\Sigmab$ that perfectly mirror its block diagonal structure. From the Davis-Kahan theorem \citep{YWS15, BD21}, we can conclude that the distance between the population and sample eigenvectors is perturbed. For the purpose of completeness, we provide the corresponding result from \citet{YWS15}.
\begin{Corollary}\label{col:DavisKahanTheorem}
    Let $\delta_i \equiv \min( \tilde{\lambda}_{i-1} - \tilde{\lambda}_i, \tilde{\lambda}_i -  \tilde{\lambda}_{i-1}) $ be the eigengap of the eigenvalue $\tilde{\lambda}_i$, where $\tilde{\lambda}_0 = \infty$ and $\tilde{\lambda}_{p+1} = -\infty$. If $\delta_i > 0$ it holds that 
    \begin{equation*}
    \Vert \tilde{\vb}_i - \vb_i \Vert \leq \dfrac{2^{3/2} \Vert \Epsilonb \Vert_{\rm op}^2 }{ \delta_i } \, ,
\end{equation*}
for $i\in\{1,\ldots,p\}$ where $\Vert \cdot \Vert_{\rm op}$ denotes the operator norm.
\end{Corollary}

In this work, we provide a concept that identifies the block diagonal structure of the population covariance matrix using sample singular vectors of the data matrix $\Xb$ also in the presence of noise.

\subsection{An Iterative Algorithm}\label{ss:Algorithm}

In Corollary~\ref{col:VectorStructure}, we conclude that the singular vectors mirror the block diagonal shape of the covariance matrix $\Sigmab$. In the sample case, however, the block diagonal structure is masked by noise. As a result, the sample singular vectors are perturbed by $\Epsilonb$ (Corollary~\ref{col:DavisKahanTheorem}) and therefore do not perfectly mirror the block diagonal structure of $\Sigmab$.


To recover their unperturbed structure and thereby reveal the block diagonal structure of the population covariance matrix, we propose the use of sparse singular vectors. These so called \textit{sparse loadings} are sparse approximations, i.e., having many zero entries, of the sample singular vectors. The intuition of using them is that by calculating sparse loadings, the zero elements of the population singular vectors, which may deviate from zero due to the perturbation $\Epsilonb$ in their estimated sample counterparts, are reset to their original zero values. Calculation of sparse loadings is a well-established area in the literature for which numerous approaches have been proposed (see, e.g., \citet{ZHT06}, \citet{SH08}, \citet{WTH09}, \citet{YMB14}, and \citet{GWS20} among others). These methods typically employ a form of regularization, such as the $\mathcal{L}_1$-type lasso constraint \citep{TI96}, the hard thresholding
penalty \citep{DJ94}, or the smoothly clipped absolute deviation penalty \citep{FL01}. We consider the former to obtain sparse loadings according to the following optimization problem:
\begin{equation}\label{eq:OptProblem}
    \min\limits_{d,\ub,\vb} \Vert \Xb - d \ub \vb^\top \Vert_F^2 \;\; {\rm subject \; to} \;\; \Vert \ub \Vert_2^2 = 1, \; \Vert \vb \Vert_2^2 = 1, \; \Vert \vb \Vert_1 \leq \alpha \, ,
\end{equation}
which calculates the first sparse loading $\check{\vb}_1$ with $\Vert \check{\vb}_1 \Vert_2^2 = 1$ and $\mathcal{L}_1$ imposing regularization parameter $\alpha$. Throughout this paper, $\Vert \vb \Vert_p$ denotes the $\mathcal{L}_p$ norm for any vector $\vb$. The remaining sparse loadings for $i>1$ can then be calculated iteratively. For this the data matrix $\Xb$ must be replaced by the residual matrices of the sequential matrix approximations.

However, the block diagonal structure can be revealed using only the first singular vector. The intuition is that the first singular vector mirrors one block which can then be omitted for further analysis. For example, assume without loss of generality that the first singular vector mirrors the first submatrix $\Xb_1$. After detection, we can proceed with the calculation of the first singular vector of the reduced data matrix $\Xb_{-1} \equiv (\Xb_2,\ldots,\Xb_b)$, omitting the first submatrix $\Xb_1$. This is done iteratively to find all submatrices one after the other.

When using sparse loadings in the practical implementation, the precision of block detection may be somewhat compromised. The first sparse loading partitions the data matrix into two submatrices: one determined by the non-zero loading components and the other by the zero-valued loading components. While these submatrices may not directly align with the actual population blocks, an recursive refinement process is employed to uncover the true underlying block structure. The steps for this recursive refinement are outlined in Algorithm~\ref{alg:BDSVD}, which illustrates the method for block detection using singular vectors (BD-SVD).

\begin{algorithm}
\caption{Procedure for BD-SVD}\label{alg:BDSVD}
\begin{algorithmic}[1]
\Statex \textbf{Input} $n \times p$ data matrix $\Xb $ with centered columns.

\vspace{0.1cm}

\State Calculate the first sparse loading $\check{\vb}_1$ with regularization parameter $\alpha$.

\vspace{0.1cm}

\State Check if $\check{\vb}_1$ mirrors one or two blocks.

\vspace{0.1cm}

\State Repeat steps 1 to 2 for the respective subsamples corresponding to the blocks detected in step 2 until the sparse loading $\check{\vb}_1$ no longer mirrors a more refined block diagonal structure. This means continuing until the first sparse loading for each subsample mirrors only one block, namely the subsample itself.

\vspace{0.1cm}

\Statex \textbf{Output} Detected structure $\Xb = (\Xb_1,\ldots,\Xb_b)$ with subsamples $\Xb_1,\ldots,\Xb_b$ of the data matrix concluding that $\Sigmab = \diag(\Sigmab_1,\ldots,\Sigmab_b)$ exhibits a block diagonal structure.
\end{algorithmic}
\end{algorithm}

\subsection{Parameter Tuning}\label{ss:ParamterTuning}

We treat the regularization $\alpha$ as tuning parameter in step 2 of Algorithm~\ref{alg:BDSVD} and the sparse loading $\check{\vb}_1 \equiv \check{\vb}_1(\alpha)$ depends on the sparsity induced by $\alpha$. Theoretically, we can increase sparsity by increasing $\alpha$ until $\check{\vb}_1$ becomes the standard base vector, suggesting that all variable are mutually uncorrelated. However, imposing such extreme sparsity may not reflect the true underlying structure of the population covariance matrix. Therefore, there is a need to control the degree of regularization and strive for a reasonable level of sparsity. In this section, we propose an approach for tuning the parameter $\alpha$.

We formulate the computation of sparse loadings as a lasso regression problem. This formulation allows us to utilize the Bayesian information criterion (BIC) \citep{SC78} for lasso regression to determine the parameter $\alpha$ that leads to the sparse loading $\check{\vb}_1 \equiv \check{\vb}_1(\alpha)$ that best fits the covariance matrix. To begin with, we note that the computation of sparse loadings can be stated as a lasso regression problem.

\begin{Remark}\label{re:SparseLoadingsToLasso}
    Assume that $\Vert \vb \Vert_2^2 = 1$. The optimization problem from (\ref{eq:OptProblem}) can be formulated as
\begin{equation*}
    \min\limits_{d, \ub, \vb} \Vert \Xb - d \ub \vb^\top \Vert_F^2 + \lambda_v  \Vert  \vb \Vert_1 \,,
\end{equation*}
where $\lambda_v\geq 0$ is a regularization parameter. For a fixed $\vb = \check{\vb}$ with $\Vert \check{\vb} \Vert_2^2 = 1$, its minimum is
\begin{equation}\label{eq:SparseLoadingAsLasso}
    \Vert \Xb - \check{\ub} \check{\vb}^\top \Vert_F^2 + \lambda_v  \Vert  \check{\vb} \Vert_1 \,,
\end{equation}
with $\check{\ub} = \Xb \check{\vb}$.
\end{Remark}

For the first sparse loading $\check{\vb}_1$, we use the optimization problem outlined above. The unit length assumption holds without loss of generality, since we can always replace the sparse loading by $\check{\vb}_1 / \Vert \check{\vb}_1  \Vert_2^2$ without affecting the sparseness of its structure. When calculating $\check{\vb}_i$ for $i>1$, the data matrix $\Xb$ needs to be replaced by the residual matrices of the sequential matrix approximations: $\Xb^{(i)} \equiv \Xb^{(i-1)} - \check{\ub}_{i-1} \check{\vb}_{i-1}^\top $ where $\check{\ub}_{i-1} = \Xb^{(i-1)}\check{\vb}_{i-1} $ and $\Xb^{(0)} \equiv \Xb$.

Now, we can adopt the BIC for lasso regression as a selection criteria. Let $\mathcal{S} \equiv \supp(\check{\vb}_1)$ be the support of the sparse loading $\check{\vb}_1$. We use $|\mathcal{S}|$ to denote the cardinality of the set $\mathcal{S}$, i.e. , the number of non-zero components in $\check{\vb}_1$. \citet{ZHT07} showed that $|\mathcal{S}|$ is an unbiased estimate for the degree of freedom of the lasso fit, and propose that the BIC can be used to select the optimal number of non-zero coefficients. We apply this result to our objective of selecting the degree of sparsity by using the connection between the calculation of sparse loadings and regularized regression (Remark~\ref{re:SparseLoadingsToLasso}). In addition, we consider a high-dimensional BIC (HBIC) which also applies in high-dimensional contexts \citep{WLL09, FT12, WKL13}.

\begin{Remark}\label{re:BICfor1}
    Let ${\rm SSR} \equiv \Vert \Xb - \check{\ub}_{1} \check{\vb}_{1}^\top \Vert_F^2 $ be the sum of squared residuals. The HBIC for the first sparse loading $\check{v}_1$ according to the lasso regularization problem (\ref{eq:SparseLoadingAsLasso}) in Remark~\ref{re:SparseLoadingsToLasso} is
\begin{equation}\label{eq:HBIC1}
{\rm HBIC} \equiv \log \left( \dfrac{{\rm SSR}}{np} \right) + |\mathcal{S}|  \dfrac{a_{np}}{np} \log(p) \, ,
\end{equation}
where $a_{np}$ is a positive sequence of numbers that diverges to infinity.
\end{Remark}


\citet{WKL13} showed that if $p_i  a_{np} \log(p) = o(np)$ for $i\in\{1,\ldots,b\}$, then HBIC identifies the true underlying model with probability approaching one under mild conditions. These conditions are also intended for scenarios in which the number of variables exceeds the number of observations in the data matrix, commonly referred to as the "$p>n$" case. However, the lasso regularization we formulated always addresses a "$p<n$" problem with $n_{\rm lasso} = np$ observations and $p_{\rm lasso} = p$ variables (online Appendix A.3), making it easier to meet the HBIC conditions. At the same time, meeting the conditions required for other criteria like the extended BIC \citep{CC08} is a greater challenge due to the constructed "$p<n$" case.

For parameter tuning, we compute $\check{\vb}_1$ for $|\mathcal{S}| \in\{0,\ldots,p-1\}$ and select the optimal sparse loading with parameter $\alpha$ according to the HBIC in (\ref{eq:HBIC1}). The choice of $a_{np} = \log(np)^{2/3}  \log\log(np)$ has proven successful in our experience.

\subsection{Illustrative Example}

In this section, we provide an illustrative example of BD-SVD. We generated a synthetic data matrix $\Xb = (\xb_1,\ldots,\xb_{3000}) = (\Xb_1, \Xb_2, \Xb_3)$ comprising $n=500$ observations and $p=3000$ variables. These variables are partitioned into three equally sized blocks, resulting in $\Xb_i$ as $500 \times 1000$ matrices for $i\in\{1,2,3\}$. The sample was drawn from an $N(\bm{0}, \Sigmab)$ distribution with population covariance matrix $\Sigmab = \diag(\Sigmab_1, \Sigmab_2, \Sigmab_3)$ with $\Sigmab_i = (1 - \omega_i) \Ib_{1000} + 2 \omega_i \bm{1}_{1000} \bm{1}_{1000}^\top$ where $\omega_i$ is uniformly $U(0.1,0.3)$ distributed and $\bm{1}_{1000}$ is a vector of ones.

\begin{figure}[!htbp]
\center
\includegraphics[width=\textwidth]{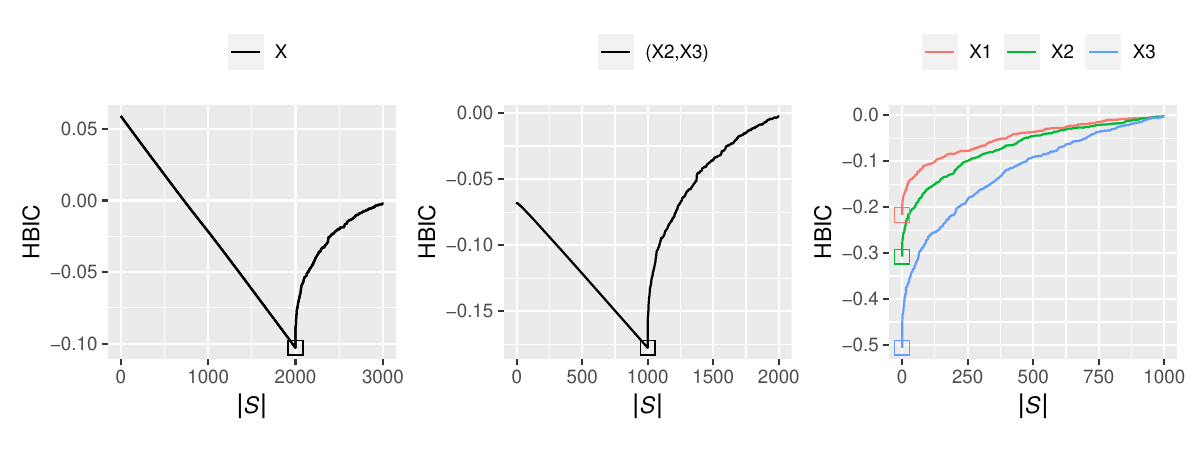}
  \caption{Illustration of block detection by BD-SVD for the data matrix $\Xb = (\Xb_1, \Xb_2, \Xb_3)$ using HBIC from (\ref{eq:HBIC1}) for parameter tuning. In the initial step \textit{(left)}, $\Xb$ is split into two blocks: $\Xb_1$ and $(\Xb_2,\Xb_3)$, where the optimal number of non-zero components is $|\mathcal{S}|=2000$ according to HBIC, and $\mathcal{S} = \{ 1001,\ldots,3000 \}$. With $|\mathcal{S}|=1000$ in the subsequent step \textit{(center)}, $(\Xb_2,\Xb_3)$ is further split into $\Xb_2$ and $\Xb_3$. The final structure has now been set. BD-SVD does not perform any further splits for $\Xb_1$, $\Xb_2$, or $\Xb_3$ \textit{(right)}, since $|\mathcal{S}|=0$ for all three matrices according to HBIC.}
  \label{fig:BICExample}
\end{figure}

Figure~\ref{fig:BICExample} illustrates the steps of BD-SVD given in Algorithm~\ref{alg:BDSVD} until all three blocks, and therefore the block diagonal structure of the population covariance matrix, are identified. For this example and throughout this work, we compute sparse loadings using the method of \citet{SH08} in step 1 of Algorithm ~\ref{alg:BDSVD}. \citet{SSM13} established consistency of this method in high-dimensional and low sample size contexts. The implementation of the method is available in the \texttt{irlba} package \citep{BRL22} within the statistical software \texttt{R} \citep{RSoftware}.

\subsection{Number of Sparse Loadings}

In Corollary~\ref{col:VectorStructure}, we conclude that the structure of a block diagonal covariance matrix is mirrored by all singular vectors. Therefore, it seems valid to choose all singular vectors ($k=p$) to identify the block diagonal structure of the covariance matrix. Further, the covariance matrix can be approximated by a reduced number of singular vectors, meaning that choosing a smaller number of vectors ($k < p$) but with $k > 1$ to achieve a good approximation could be a good choice as well. The advantage is that we avoid calculating the first sparse loading $\check{\vb}_1$ recursively until the block diagonal structure of the covariance is revealed, but rather calculate $\check{\Vb} = (\check{\vb}_1,\ldots,\check{\vb}_k)$ with $k>1$ that mirrors the true shape directly.

We recap (see, e.g., \citet{EY36}) that for any $k\leq r$
\begin{equation*}
    \Sb^{(k)} \equiv \sum\limits_{i \in \{1,\ldots,k\}} \lambda_i \bm{v}_i \bm{v}_i^\top = \argmin\limits_{\hat{\Sb} \,:\, \rank(\hat{\Sb})\leq k} \Vert \Sb - \hat{\Sb} \Vert_F^2 \; ,
\end{equation*}
is the best rank-$k$ approximation to $\Sb$ in the sense of the squared Frobenius norm ($\Vert \cdot \Vert_F^2$). If we are able to reveal the block diagonal structure from $\Sb$ using all singular vectors (Corollary~\ref{col:VectorStructure}), it is reasonable that this is also possible using $k$ singular vectors of a low rank approximation $\Sb^{(k)}$ of $\Sb$ with $k<r$.

In Remark~\ref{re:SparseLoadingsToLasso}, we recapped calculation of $\check{\vb}_i$ for $i\geq 1$. Tuning of the regularization parameter $\alpha$ using the HBIC can be done as follows.

\begin{Remark}\label{re:BICforK}
    Let ${\rm SSR}(k) \equiv \sum_{i \in \{1,\ldots,k\}} \Vert \Xb^{(i)} - \check{\ub}_{i} \check{\vb}_{i}^\top \Vert_F^2 $ and let $|\mathcal{S}_i|$ be the support of $\check{\vb}_i$ such that $|\mathcal{S}(k)| \equiv \sum_{i\in\{1,\ldots,k\}} |\mathcal{S}_i|$ denotes the overall degree of sparsity. The HBIC for the first $k$ sparse loadings according to the lasso regularization problem in Remark~\ref{re:SparseLoadingsToLasso} is
\begin{equation*}
{\rm HBIC} \equiv \log \left( \dfrac{{\rm SSR}(k)}{npk} \right) + |\mathcal{S}(k)| \dfrac{a_{npk}}{npk} \log(pk)  \,.
\end{equation*}
\end{Remark}

Singular vectors corresponding to singular values with multiplicity greater than one, or those corresponding to singular values that are nearly equal to each other, are strongly masked in the sample case (Corollary~\ref{col:DavisKahanTheorem}). This poses a challenge in revealing their inherent structure and argues in favor of not choosing singular vectors corresponding to these singular values. With $k=1$ however, this concern is less likely to occur. In particular, spiked covariance models \citep{PA07, JL09} which are frequently used to model high-dimensional phenomena, indicate that singular values outside the spikes are in close distance, which also argues in favor of not using all singular vectors.

\section{Simulation Studies}\label{s:Simulations}

In this section, we evaluate the performance of BD-SVD described in Section~\ref{s:BDSVD} with $k=1$. We simulate a data matrix containing $n$ observations from a $p$-multivariate normal distribution $N(\bm{0}, \Sigmab)$. BD-SVD is compared to three other approaches:

\begin{enumerate}
    \item SHDJ and SHRR: In their recent work, \citet{DG18} demonstrated that their methods (SHDJ and SHRR) outperformed existing ones. Therefore, these methods can be considered to be state of the art and we therefore compare BD-SVD to them. The methods are implemented in the \texttt{R} package \texttt{shock} \citep{DG15}.

    \item Est$_\tau$: An \textit{ad hoc} method for detecting the block diagonal structure within the covariance matrix involves estimating the sample covariance matrix and applying hard thresholding. In this method, all absolute values below a certain threshold $\tau$ are set to zero. Several high-dimensional covariance matrix estimation methods exist (see, e.g., \citet{BL08}, \citet{RLZ09}, \citet{FLM13}, among others). Estimation is performed by generalized thresholding of the covariance matrix \citep{FL01, RLZ09} available in the statistical software \texttt{R} by \citet{BHCvLD21}, with $\tau$ values chosen from $\{0.1, 0.2\}$.

    \item SPCA: Principal component analysis is a widely used technique for dimensionality reduction and interpretation of a data matrix \citep{JWTT21}. It generates principal components $\vb_1\Xb, \ldots, \vb_p \Xb$ using the singular vectors $\vb_1, \ldots, \vb_p$, where  the variance of each principal component is its corresponding eigenvalue $\lambda_i$. \\
    However, interpreting the principal components, which are linear combinations of all variables, can be challenging. Sparse principal component analysis (SPCA) overcomes this disadvantage by setting some components of the singular vectors to zero, which improves interpretability at the price of a lower explained variance. Typically, calculation of these sparse loadings is based on solving the optimization problem in (\ref{eq:OptProblem}). Notably, the methods mentioned in Section~\ref{ss:Algorithm} to calculate sparse loadings for our block detection objective were originally developed in the context of SPCA.
\end{enumerate}

We evaluate the performance using sensitivity (Sensitivity = TP/(TP + FN)), specificity (Specificity = TN/(TN + FP)), and the false discovery rate (FDR) (FDR = FP/(TP + FP)). Here for $i\in\{1,\ldots,b\}$, TP is the number of true positive detection (separating $\Sigmab_i$ from the other blocks), FP is the number of false positive  detection (splitting $\Sigmab_i$ into smaller blocks), TN is the number of true negative detection (not splitting $\Sigmab_i$ into smaller blocks), and FN is the number of false negative detection (not separating $\Sigmab_i$ from the other blocks)

All simulation results were obtained using the statistical software \texttt{R} 4.2.1 on a computational cluster running Rocky Linux 8.

\subsection{Diagonal Covariance Structure}
We examine scenarios in which the data matrix consists of mutually uncorrelated variables ($p=b)$. Specifically, we consider two cases: (a) $\Sigmab = \Ib_p$ and (b) $\Sigmab = \diag(\sigma_1, \ldots, \sigma_p)$, where $\sigma_i$ is uniformly $U(1,5)$ distributed. Across both simulation designs, we use $(n,p) \in \{(65,500), 
 \allowbreak (125, 500), \allowbreak (250,500)\}$.

\begin{figure}[!htbp]
\centering
\begin{subfigure}{1\textwidth}
    \includegraphics[width=\textwidth]{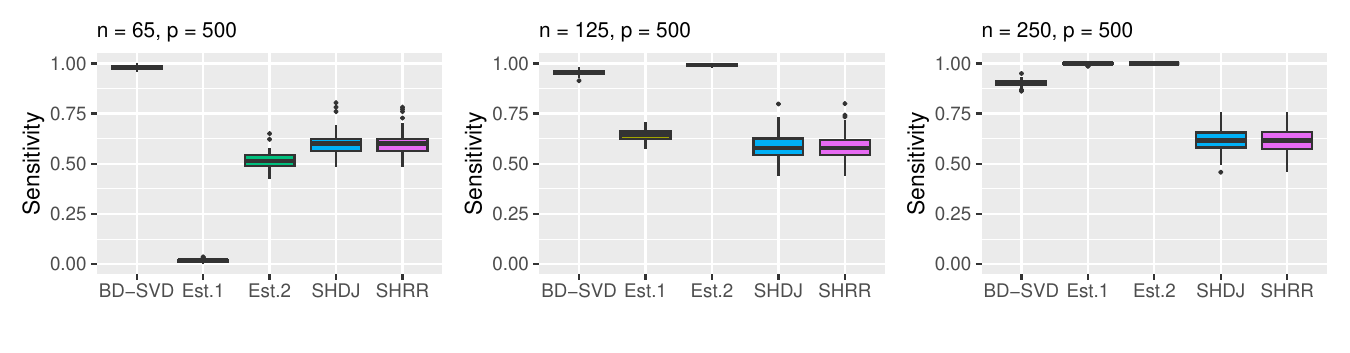}
    \caption{Simulation design (a).}
\end{subfigure}
\hfill
\begin{subfigure}{1\textwidth} 
    \includegraphics[width=\textwidth]{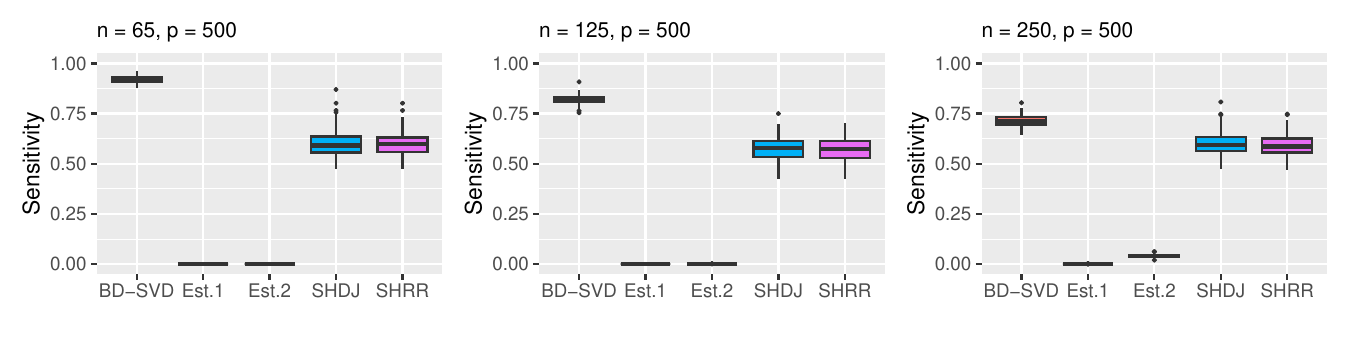}
    \caption{Simulation design (b).}
\end{subfigure}
\caption{Simulation designs (a) and (b): Performance of BD-SVD, SHDJ, and SHRR for  $(n,p) \in \{(65,500), 
 \allowbreak (125, 500), \allowbreak (250,500)\}$ measured by Sensitivity.}
\label{fig:SimulationAB}
\end{figure}

The results for both simulation designs are illustrated in Figure~\ref{fig:SimulationAB}. Since there can be no false positive block detection for diagonal matrices, Specificity = 1 and FDR = 0 for these designs. Consequently, only the sensitivity is given.

\subsection{Block Diagonal Covariance Structure}
In this section, we examine scenarios in which the covariance matrix exhibits a block diagonal structure $\Sigmab = \diag(\Sigmab_1,\ldots,\Sigmab_b)$ with $b=10$ blocks $\Sigmab_i$ of equal sizes. Specifically, we consider two cases: (c) each block exhibits a compound symmetric covariance structure $\Sigmab_i = (1 - \omega_i) \Ib_{p_i} + 2 \omega_i \bm{1}_{p_i} \bm{1}_{p_i}^\top$ where $\omega_i$ is uniformly $U(0.1,0.3)$ distributed and $\bm{1}_{p_i}$ is a vector of ones, and (d) $\Sigmab_i = \bm{M}(i)$ where the components of $\bm{M}(l) = (m_{ij})$, simulated as $m_{ij} = (-1)^{i+j}  \omega_l^{|i-j|^{0.05}}$ with $\omega_l \sim U(0.3,0.5)$, decrease in magnitude the further they move away from the diagonal. Across both simulation designs, we consider $(n,p) \in \{(250, 500),\allowbreak (500,500), \allowbreak (500, 2500), \allowbreak (500, 5000)\}$ such that all blocks $\Sigmab_i$ are of equal size $p_i = p/b \in \{50,\allowbreak 250,\allowbreak 500\}$.

\begin{figure}[!htbp]
\center
\includegraphics[width=1\textwidth]{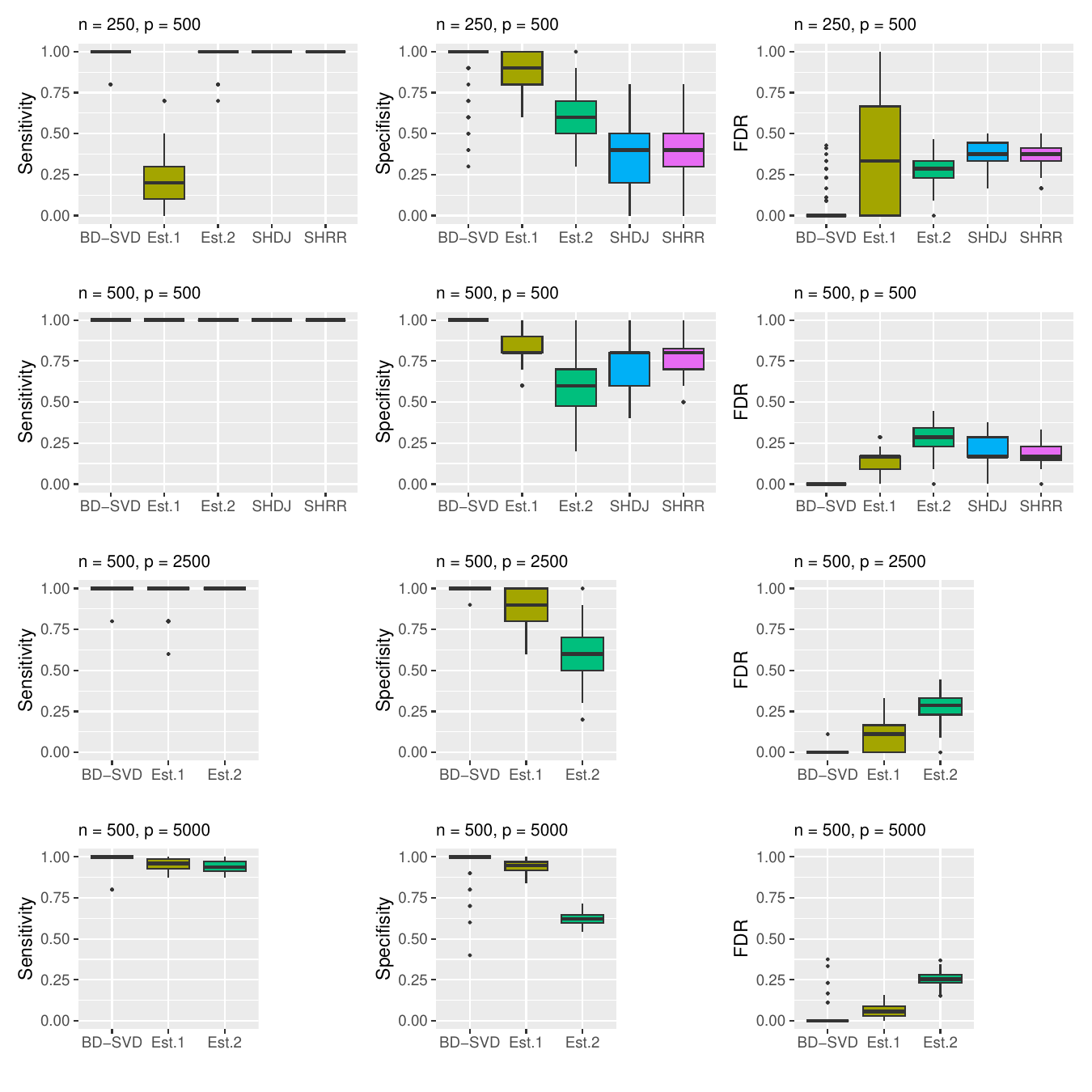}
  \caption{Simulation design (c): Performance of BD-SVD, Est$_{0.1}$, Est$_{0.2}$, SHDJ, and SHRR for $(n,p) \in \{(250, 500),\allowbreak (500, 500)\}$, and of BD-SVD, Est$_{0.1}$, and Est$_{0.2}$ for $(n,p) \in \{(500, 2500), \allowbreak (500, 5000)\}$ measured by Sensitivity, Specificity, and FDR.}
  \label{fig:SimulationC}
\end{figure}

\begin{figure}[!htbp]
\center
\includegraphics[width=1\textwidth]{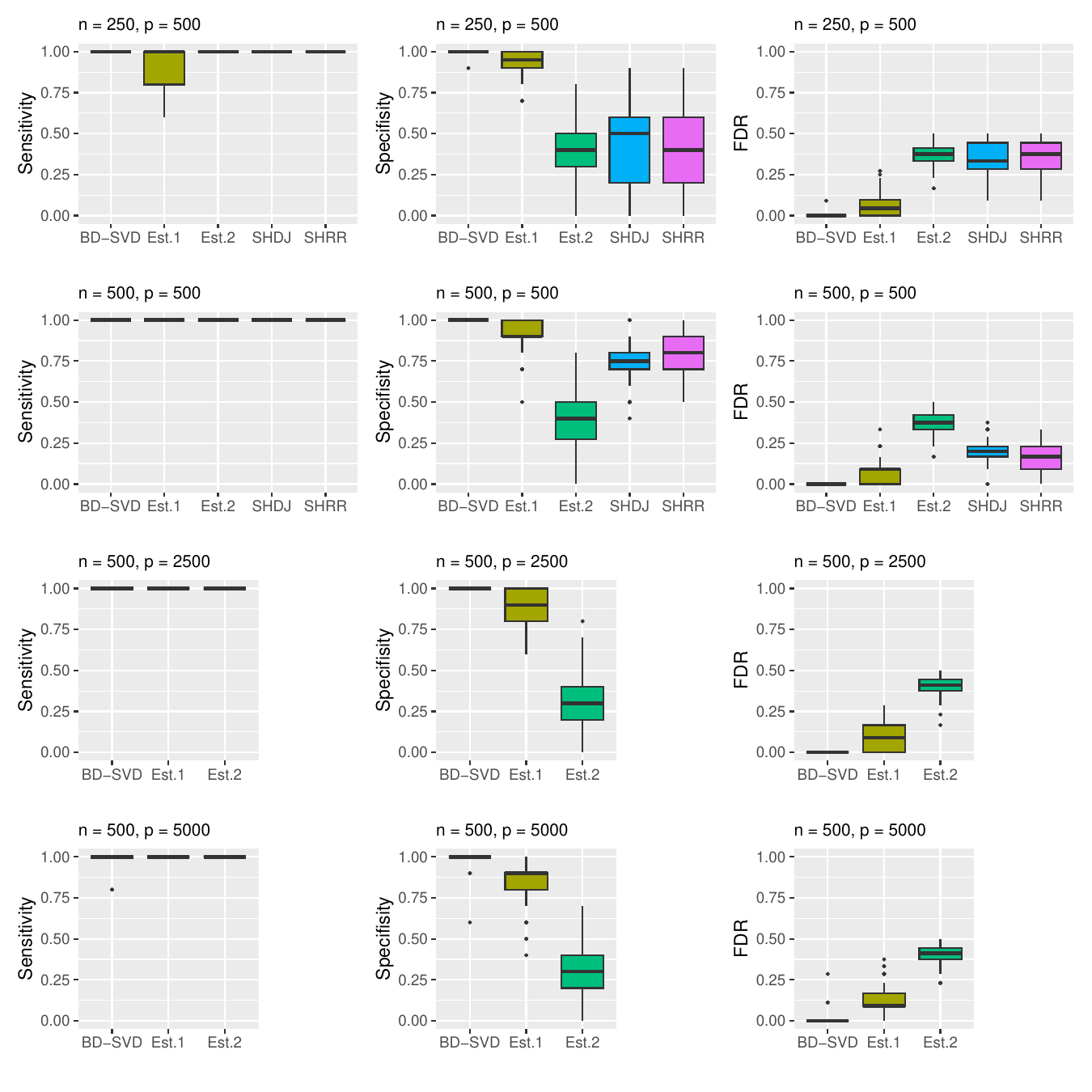}
  \caption{Simulation design (d): Performance of BD-SVD, Est$_{0.1}$, Est$_{0.2}$, SHDJ, and SHRR for $(n,p) \in \{(250, 500),\allowbreak (500, 500)\}$, and of BD-SVD, Est$_{0.1}$, and Est$_{0.2}$ for $(n,p) \in \{(500, 2500), \allowbreak (500, 5000)\}$ measured by Sensitivity, Specificity, and FDR.}
  \label{fig:SimulationD}
\end{figure}

The results for simulation designs (c) and (d) are illustrated in Figure~\ref{fig:SimulationC} and Figure~\ref{fig:SimulationD} respectively. Notably, the comparison of BD-SVD with SHDJ and SHRR was limited to cases where $p \leq 500$ due to the long running time of these methods in higher dimensions. In fact, the methods did not converge in our simulations when $p \geq 2500$.

BD-SVD outperforms the other methods across the simulation designs. It is worth noting that the \textit{ad hoc} procedure shows decent performance, although not reliably across all designs. Intuitively, smaller values for $\tau$ lead to a higher sensitivity while large vales lead to a higher specificity.

\subsection{A Note on Sparse Principal Component Analysis}

 Although both BD-SVD and SPCA employ sparse loadings to uncover data structures, they differ in their goals. While SPCA aims to identify sparse loadings explaining the most variance in the data matrix, BD-SVD targets sparse loadings fitting the population covariance matrix. Though their aims may occasionally align, they typically diverge.

To illustrate this, we ran SPCA on simulation design (c) and (d) with $n=500$, $p=500$, and $b=10$ such that $p_i=50$. Simplifying the process, we computed only the first SPCA loading. To determine the suitable sparseness for SPCA, we increased sparsity until the first sparse principal component explained at least 90\% or 95\% of the first principal component without sparseness constraints, a common practice in SPCA \citep{ZHT06, SH08}.

\begin{table}[]
    \centering
\begin{tabular}{SSSSSSS} \toprule
& {Expl.} & {$n$} & {$p$} & {$b$} & {\textbf{Sensitivity}}\\ \midrule
{Simulation design (c)} & 0.90  & 500   & 500   & 10  & \textbf{0.47}   \\
{Simulation design (c)} & 0.95  & 500   & 500   & 10  & \textbf{0.67}   \\
{Simulation design (d)} & 0.90  & 500   & 500   & 10  & \textbf{0.60}   \\
{Simulation design (d)} & 0.95  & 500   & 500   & 10  & \textbf{0.70}   \\
\bottomrule
\end{tabular}
    \caption{Sensitivity of the first sparse sparse principal component for simulation designs (c) and (d) Sparseness for SPCA was increased until the first sparse principal component explained at least $90\%$ or $95\%$ of the first principal component without sparseness constraints.}
    \label{tab:SimulationSCPA}
\end{table}

Table~\ref{tab:SimulationSCPA} contains the sensitivity analysis of the simulation study, indicating that SCPA is less effective than BD-SVD. It is important to emphasize that the study only evaluates whether the first sparse principal component reflects the block diagonal structure. To reveal the entire structure, SPCA must be continued iteratively for the detected blocks, which leads to a further decrease in sensitivity. Additionally, in this simulation study, we calculated the explained variance with respect to the known population eigenvalue. However, in high-dimensional contexts, estimating the eigenvalue poses challenges (see, e.g., \citet{JO01}, \citet{BS06}, and \citet{PA07}, among others), which impacts the performance of SPCA.

\section{Real Data Examples}\label{s:Examples}

\subsection{Lung Cancer Data}\label{ss:LungExample}
In this section, we illustrate BD-SVD using microarray gene expression signatures. Specifically, we analyze a lung cancer gene expression data set that consists of $203$ patient samples comprising 139 lung adenocarcinomas, 21 squamous cell carcinoma cases, 20 pulmonary carcinoid tumors, 6 small cell lung cancers, and 17 normal lung samples. The data and additional details can be found online in the supporting information section of the article by \citet{BH01}.

The  original  data set contains  $12\,600$ gene expressions measured using the Affymetrix 95av2 GeneChip.  Following procedures similar to \citet{LHNM08} and \citet{RLZ09}, we filter the genes using the ratio of the sample standard deviation and sample mean of each gene. We keep the $600$ genes with the highest ratio and the $600$ genes with the lowest ratio, and refer to them as \textit{high-ratio genes} and \textit{low-ratio genes} respectively. We then standardize the remaining genes so that each gene has a sample mean of 0 and a sample standard deviation of 1. After gene filtering, the data set contains $n=203$ patients with $p=1200$ genes.

\begin{figure}[!htbp]
\center
\includegraphics[width=0.8\textwidth]{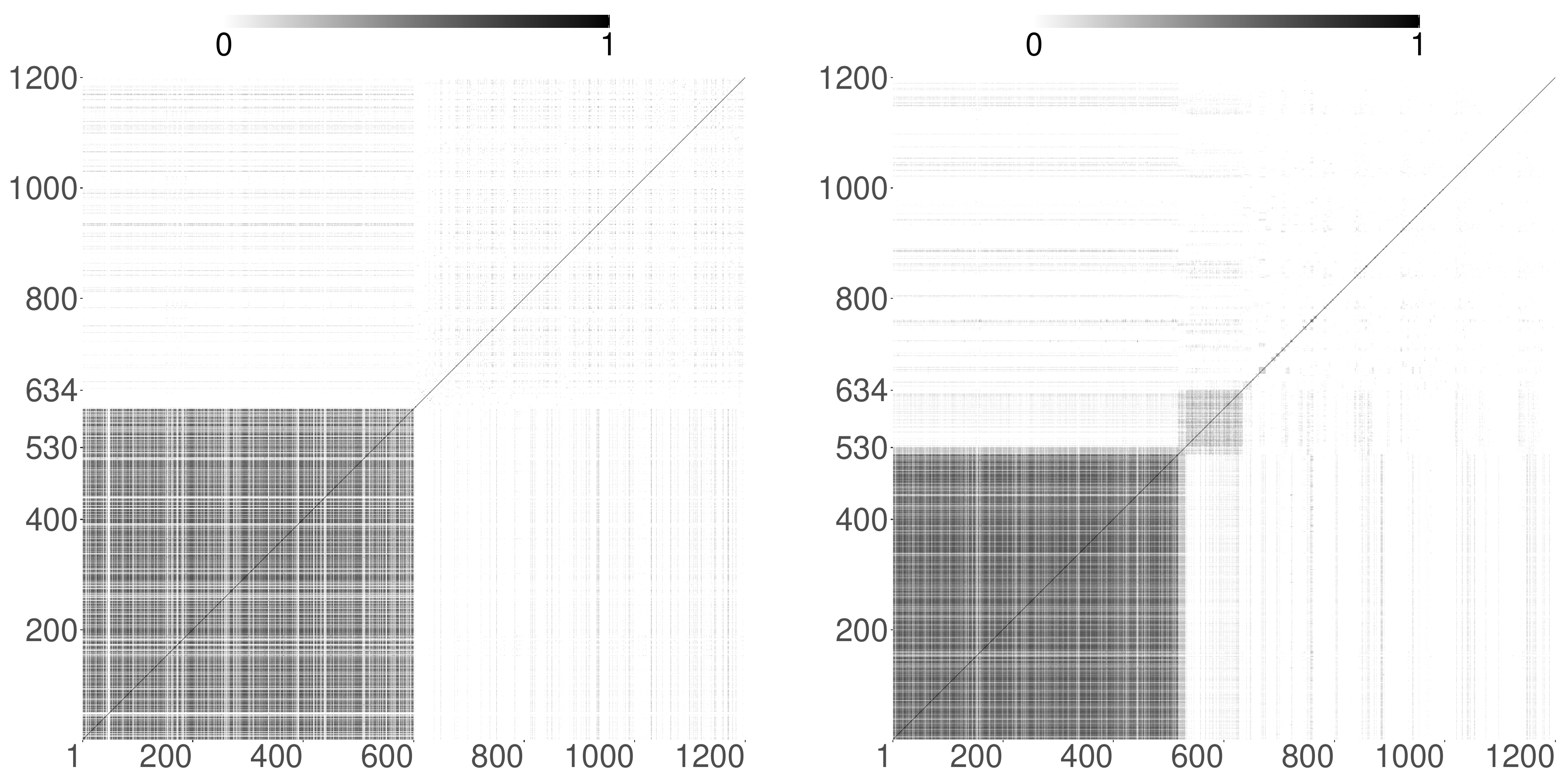}
  \caption{Sample covariance matrix with absolute values for the standardized gene expressions of the lung cancer data in its original data-ordering (\textit{left}) and after reordering the variables based on the detected blocks $\Sigmab_1,\ldots,\Sigmab_{217}$ (\textit{right}).}
  \label{fig:LungCancerExample}
\end{figure}

The left plot in Figure~\ref{fig:LungCancerExample} illustrates the sample covariance matrix of this subsample. We use absolute values of the estimated correlation coefficients because we are interested in the strength of the pairwise association between the genes, regardless of their sign. BD-SVD identifies 217 blocks $\Sigmab_1, \ldots, \Sigmab_{217}$ with varying block sizes $p_1 = 530$, $p_2 = 104$, and $p_i \leq 3$ for $i\in\{3,\ldots,217\}$. Figure~\ref{fig:LungCancerExample} \textit{(right)} illustrates the sample covariance matrix after permuting variables according to the identified blocks.

A visual analysis of the first 600 variables in the left-hand plot, which correspond to the high-ratio genes, initially suggests a strong correlation structure, with little correlation between the low-ratio genes. However, BD-SVD detects a more refined structure. For example, the two largest blocks $\Sigmab_1$ and $\Sigmab_2$ contain a total of $p_1 + p_2 = 634$ variables. Notably, $97.55\%$ of the variables in $\Sigmab_1$ and $81.55\%$ of the variables in $\Sigmab_2$ are high-ratio genes. On the other hand, $13.83\%$ of the variables contained in the remaining blocks $\Sigmab_i$ with $i \in\{3,\ldots,217\}$, i.e., all blocks except $\Sigmab_1$ and $\Sigmab_2$, are high-ratio genes. Thus, not all high-ratio genes are correlated with each other: Some high-ratio genes show no correlation with other high-ratio genes, while there are also cases in which low-ratio genes correlate with high-ratio genes.

\subsection{Daily Stock Returns}\label{ss:StockExample}

In this section, we consider cross-sectional correlation of daily stock returns from various sectors from the S\&P 500, including mining, quarrying, and oil and gas extraction sector; utilities sector; wholesale trade sector; retail trade sector; transportation and warehousing sector; information sector; finance and insurance sector; real estate and rental and leasing sector; professional, scientific, and technical services sector; administrative and support and waste management and remediation services sector; and arts, entertainment, and recreation sector. The data comes from the Center for Research in Security Prices and is available through Wharton Research Data Services. It consists of closing prices or bid/ask averages of 278 stocks on the trading days in the last quarter of 2022, which ranges from October 1, 2022 to December 31, 2022, encompassing a total of 63 days. Consequently, the data matrix consists of $n=63$ observations and $p=278$ variables.

\begin{figure}[!htbp]
\center
\includegraphics[width=0.8\textwidth]{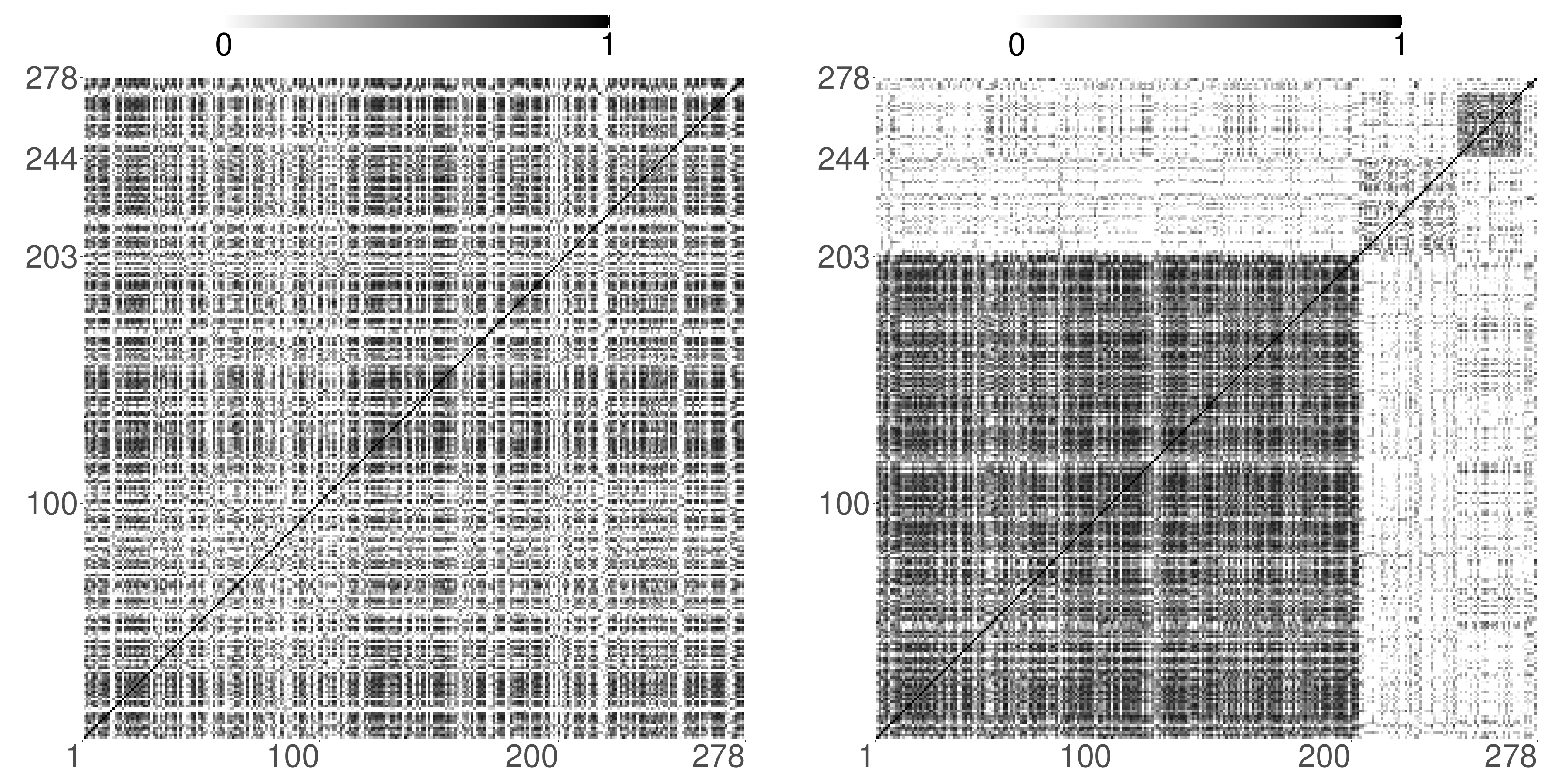}
  \caption{Sample covariance matrix with absolute values for the standardized daily stock returns data in its original data-ordering (\textit{left}) and after reordering the variables based on the detected blocks $\Sigmab_1,\ldots,\Sigmab_5$ (\textit{right}).}
  \label{fig:SPExample}
\end{figure}

We prepare the data matrix by standardizing it prior to our analysis. BD-SVD detects five blocks $\Sigmab_1,\ldots,\Sigmab_5$ with $p_1 = 203$, $p_2 = 41$, $p_3=27$, $p_4 = 5$, and $p_5 = 2$. The sample covariance matrix, with variables permuted according to our findings, is illustrated in Figure~\ref{fig:SPExample}. We use the absolute values of the estimated correlation coefficients for the reasons explained in the previous example.

\section{Discussion}\label{s:Discussion}

In this paper, we present a nonparametric method for detecting block diagonal structures in covariance matrices within high-dimensional data. Our approach relies on the first sparse loading ($k = 1$) to mirror the underlying covariance matrix structure. Although we acknowledge the potential usefulness of employing $k > 1$ sparse loadings, we intentionally limit our exploration of this scenario. This decision is motivated by the considerable accuracy achieved with the first loading only, coupled with potential drawbacks associated with using additional loadings. These drawbacks include increased computational costs and the risk of a stronger masking effect on the true structure of the covariance matrix. However, in our experience, these measures often resulted in incorrect covariance structures.

The choice of the appropriate degree of sparsity for the singular vectors plays a fundamental role in our methodology. To address this, we suggest to formulate the computation of the sparse loadings as a lasso regression problem. This allows us to use the HBIC to determine the appropriate sparsity level.

For completeness, we want to mention that in the course of our research we also experimented with other approaches to determine the appropriate degree of sparsity. For example, we tried cross validation to select the appropriate sparseness in our formulated lasso regression problem. However, as in the findings of \citet{WLT07}, this approach resulted in loadings that were not sufficiently sparse. To evaluate the validity of the detected blocks, we also considered dependency measures. In high-dimensional contexts, several methods have been developed to measure dependencies among subvectors (see, e.g., \citet{SR13}, \citet{SPV20}, \citet{PWZZZ20}, \citet{ZZYS20}, and \citet{CH21} among others). These measures typically yield zero values when the correlations between subvectors are zero. Furthermore, for our purposes, it would be necessary to establish a threshold value to determine when blocks can be considered uncorrelated.


BD-SVD has been implemented in the newly developed \texttt{R} package \texttt{bdsvd} \textit{[reference not given to maintaining blind review]}, which was written specifically for this research project.

\bigskip
\begin{center}
{\large\bf SUPPLEMENTARY MATERIAL}
\end{center}

The online supplementary materials contain derivations and proofs, and \texttt{R} code to perform the illustrative example and the simulation studies.




\bibliographystyle{chicago}

\bibliography{Bibliography}

\begin{thebibliography}{}

\bibitem[\protect\citeauthoryear{Baglama, Reichel, and Lewis}{Baglama
  et~al.}{2022}]{BRL22}
Baglama, J., L.~Reichel, and B.~W. Lewis (2022).
\newblock {\em irlba: Fast Truncated Singular Value Decomposition and Principal
  Components Analysis for Large Dense and Sparse Matrices}.
\newblock R package version 2.3.5.1.

\bibitem[\protect\citeauthoryear{Bai, Jiang, Yao, and Zheng}{Bai
  et~al.}{2009}]{BJYZ09}
Bai, Z., D.~Jiang, J.-F. Yao, and S.~Zheng (2009).
\newblock {Corrections to LRT on large-dimensional covariance matrix by RMT}.
\newblock {\em Ann. Stat.\/}~{\em 37\/}(6B), 3822--3840.

\bibitem[\protect\citeauthoryear{Baik and Silverstein}{Baik and
  Silverstein}{2006}]{BS06}
Baik, J. and J.~W. Silverstein (2006).
\newblock Eigenvalues of large sample covariance matrices of spiked population
  models.
\newblock {\em J. Multivar. Anal.\/}~{\em 97\/}(6), 1382--1408.

\bibitem[\protect\citeauthoryear{Bao, Hu, Pan, and Zhou}{Bao
  et~al.}{2017}]{BHPZ17}
Bao, Z., J.~Hu, G.~Pan, and W.~Zhou (2017).
\newblock {Test of independence for high-dimensional random vectors based on
  freeness in block correlation matrices}.
\newblock {\em Electron. J. Stat.\/}~{\em 11\/}(1), 1527--1548.

\bibitem[\protect\citeauthoryear{Bauer and Drabant}{Bauer and
  Drabant}{2021}]{BD21}
Bauer, J.~O. and B.~Drabant (2021).
\newblock {Principal loading analysis}.
\newblock {\em {J. Multivar. Anal.}\/}~{\em 184}.

\bibitem[\protect\citeauthoryear{Bhattacharjee, Richards, Staunton, Li, Monti,
  Vasa, Ladd, Beheshti, Bueno, Gillette, Loda, Weber, Mark, Lander, Wong,
  Johnson, Golub, Sugarbaker, and Meyerson}{Bhattacharjee et~al.}{2001}]{BH01}
Bhattacharjee, A., W.~G. Richards, J.~E. Staunton, C.~Li, S.~Monti, P.~P. Vasa,
  C.~Ladd, J.~Beheshti, R.~Bueno, M.~A. Gillette, M.~Loda, G.~Weber, E.~J.
  Mark, E.~S. Lander, W.~Wong, B.~E. Johnson, T.~R. Golub, D.~J. Sugarbaker,
  and M.~L. Meyerson (2001).
\newblock Classification of human lung carcinomas by {mRNA} expression
  profiling reveals distinct adenocarcinoma subclasses.
\newblock {\em Proc. Natl. Acad. Sci. U.S.A.\/}~{\em 98}, 13790--13795.

\bibitem[\protect\citeauthoryear{Bickel and Levina}{Bickel and
  Levina}{2008}]{BL08}
Bickel, P.~J. and E.~Levina (2008).
\newblock Covariance regularization by thresholding.
\newblock {\em Ann. Stat.\/}~{\em 36\/}(6), 2577--2604.

\bibitem[\protect\citeauthoryear{Birke and Dette}{Birke and Dette}{2005}]{BD05}
Birke, M. and H.~Dette (2005).
\newblock A note on testing the covariance matrix for large dimension.
\newblock {\em Statist. Probab. Lett.\/}~{\em 74\/}(3), 281--289.

\bibitem[\protect\citeauthoryear{Boileau, Hejazi, Collica, van~der Laan, and
  Dudoit}{Boileau et~al.}{2021}]{BHCvLD21}
Boileau, P., N.~S. Hejazi, B.~Collica, M.~J. van~der Laan, and S.~Dudoit
  (2021).
\newblock {cvCovEst}: Cross-validated covariance matrix estimator selection and
  evaluation in {R}.
\newblock {\em J. Open Source Softw.\/}~{\em 6\/}(63), 3273.

\bibitem[\protect\citeauthoryear{Chatterjee}{Chatterjee}{2021}]{CH21}
Chatterjee, S. (2021).
\newblock A new coefficient of correlation.
\newblock {\em J. Am. Stat. Assoc.\/}~{\em 116\/}(536), 2009--2022.

\bibitem[\protect\citeauthoryear{Chen and Chen}{Chen and Chen}{2008}]{CC08}
Chen, J. and Z.~Chen (2008, 09).
\newblock {Extended Bayesian information criteria for model selection with
  large model spaces}.
\newblock {\em Biometrika\/}~{\em 95\/}(3), 759--771.

\bibitem[\protect\citeauthoryear{Chen, Zhang, and Zhong}{Chen
  et~al.}{2010}]{CZZ10}
Chen, S.~X., L.-X. Zhang, and P.-S. Zhong (2010).
\newblock Tests for high-dimensional covariance matrices.
\newblock {\em J. Am. Stat. Assoc.\/}~{\em 105\/}(490), 810--819.

\bibitem[\protect\citeauthoryear{Devijver and Gallopin}{Devijver and
  Gallopin}{2015}]{DG15}
Devijver, E. and M.~Gallopin (2015).
\newblock {\em shock: Slope Heuristic for Block-Diagonal Covariance Selection
  in High Dimensional Gaussian Graphical Models}.
\newblock R package version 1.0.

\bibitem[\protect\citeauthoryear{Devijver and Gallopin}{Devijver and
  Gallopin}{2018}]{DG18}
Devijver, E. and M.~Gallopin (2018).
\newblock Block-diagonal covariance selection for high-dimensional gaussian
  graphical models.
\newblock {\em J. Am. Stat. Assoc.\/}~{\em 113\/}(521), 306--314.

\bibitem[\protect\citeauthoryear{Donoho and Johnstone}{Donoho and
  Johnstone}{1994}]{DJ94}
Donoho, D.~L. and I.~M. Johnstone (1994).
\newblock Ideal spatial adaptation by wavelet shrinkage.
\newblock {\em Biometrika\/}~{\em 81\/}(3), 425--455.

\bibitem[\protect\citeauthoryear{Eckart and Young}{Eckart and
  Young}{1936}]{EY36}
Eckart, C. and G.~Young (1936).
\newblock The approximation of one matrix by another of lower rank.
\newblock {\em Psychometrika\/}~{\em 1}, 211–218.

\bibitem[\protect\citeauthoryear{Fan and Li}{Fan and Li}{2001}]{FL01}
Fan, J. and R.~Li (2001).
\newblock Variable selection via nonconcave penalized likelihood and its oracle
  properties.
\newblock {\em J. Am. Stat. Assoc.\/}~{\em 96\/}(456), 1348--1360.

\bibitem[\protect\citeauthoryear{Fan and Li}{Fan and Li}{2006}]{FL06}
Fan, J. and R.~Li (2006).
\newblock Statistical challenges with high dimensionality: Feature selection in
  knowledge discovery.
\newblock In {\em Proceedings of the International Congress of Mathematicians},
  Volume~3, pp.\  595--622. European Mathematical Society.

\bibitem[\protect\citeauthoryear{Fan, Liao, and Mincheva}{Fan
  et~al.}{2013}]{FLM13}
Fan, J., Y.~Liao, and M.~Mincheva (2013).
\newblock Large covariance estimation by thresholding principal orthogonal
  complements.
\newblock {\em J. R. Stat. Soc. B\/}~{\em 75\/}(4), 603--680.

\bibitem[\protect\citeauthoryear{Fan and Tang}{Fan and Tang}{2012}]{FT12}
Fan, Y. and C.~Y. Tang (2012, 12).
\newblock Tuning parameter selection in high dimensional penalized likelihood.
\newblock {\em J. R. Stat. Soc. B\/}~{\em 75\/}(3), 531--552.

\bibitem[\protect\citeauthoryear{Fisher, Sun, and Gallagher}{Fisher
  et~al.}{2010}]{FSG10}
Fisher, T.~J., X.~Sun, and C.~M. Gallagher (2010).
\newblock A new test for sphericity of the covariance matrix for high
  dimensional data.
\newblock {\em {J. Multivar. Anal.}\/}~{\em 101\/}(10), 2554--2570.

\bibitem[\protect\citeauthoryear{Friedman, Hastie, and Tibshirani}{Friedman
  et~al.}{2007}]{FHT07}
Friedman, J., T.~Hastie, and R.~Tibshirani (2007, 12).
\newblock {Sparse inverse covariance estimation with the graphical lasso}.
\newblock {\em Biostatistics\/}~{\em 9\/}(3), 432--441.

\bibitem[\protect\citeauthoryear{Gataric, Wang, and Samworth}{Gataric
  et~al.}{2020}]{GWS20}
Gataric, M., T.~Wang, and R.~J. Samworth (2020).
\newblock Sparse principal component analysis via axis-aligned random
  projections.
\newblock {\em J. R. Stat. Soc. B\/}~{\em 82\/}(2), 329--359.

\bibitem[\protect\citeauthoryear{Hyodo, Shutoh, Nishiyama, and Pavlenko}{Hyodo
  et~al.}{2015}]{HSNP15}
Hyodo, M., N.~Shutoh, T.~Nishiyama, and T.~Pavlenko (2015).
\newblock Testing block-diagonal covariance structure for high-dimensional
  data.
\newblock {\em Stat. Neerl.\/}~{\em 69\/}(4), 460--482.

\bibitem[\protect\citeauthoryear{James, Witten, Hastie, and Tibshirani}{James
  et~al.}{2021}]{JWTT21}
James, G., D.~Witten, T.~Hastie, and R.~Tibshirani (2021).
\newblock {\em An Introduction to Statistical Learning with Applications in
  R\/} (2nd ed.).
\newblock New York, USA: Springer.

\bibitem[\protect\citeauthoryear{Jiang, Bai, and Zheng}{Jiang
  et~al.}{2013}]{JBZ13}
Jiang, D., Z.~Bai, and S.~Zheng (2013).
\newblock Testing the independence of sets of large-dimensional variables.
\newblock {\em Sci. China Math.\/}~{\em 56}, 135--147.

\bibitem[\protect\citeauthoryear{Jiang and Yang}{Jiang and Yang}{2013}]{JY13}
Jiang, T. and F.~Yang (2013).
\newblock {Central limit theorems for classical likelihood ratio tests for
  high-dimensional normal distributions}.
\newblock {\em Ann. Stat.\/}~{\em 41\/}(4), 2029--2074.

\bibitem[\protect\citeauthoryear{Johnstone}{Johnstone}{2001}]{JO01}
Johnstone, I.~M. (2001).
\newblock {On the distribution of the largest eigenvalue in principal
  components analysis}.
\newblock {\em Ann. Stat.\/}~{\em 29\/}(2), 295--327.

\bibitem[\protect\citeauthoryear{Johnstone}{Johnstone}{2008}]{JO08}
Johnstone, I.~M. (2008).
\newblock {Multivariate analysis and Jacobi ensembles: Largest eigenvalue,
  Tracy–Widom limits and rates of convergence}.
\newblock {\em Ann. Stat.\/}~{\em 36\/}(6), 2638--2716.

\bibitem[\protect\citeauthoryear{Johnstone and Lu}{Johnstone and
  Lu}{2009}]{JL09}
Johnstone, I.~M. and A.~Y. Lu (2009).
\newblock On consistency and sparsity for principal components analysis in high
  dimensions.
\newblock {\em J. Am. Stat. Assoc.\/}~{\em 104\/}(486), 682--693.

\bibitem[\protect\citeauthoryear{Ledoit and Wolf}{Ledoit and Wolf}{2002}]{LW02}
Ledoit, O. and M.~Wolf (2002).
\newblock {Some hypothesis tests for the covariance matrix when the dimension
  is large compared to the sample size}.
\newblock {\em Ann. Stat.\/}~{\em 30\/}(4), 1081--1102.

\bibitem[\protect\citeauthoryear{Leung and Drton}{Leung and Drton}{2018}]{LD18}
Leung, D. and M.~Drton (2018).
\newblock Testing independence in high dimensions with sums of rank
  correlations.
\newblock {\em Ann. Stat.\/}~{\em 46\/}(1), 280--307.

\bibitem[\protect\citeauthoryear{Liu, Hayes, Nobel, and Marron}{Liu
  et~al.}{2008}]{LHNM08}
Liu, Y., D.~N. Hayes, A.~Nobel, and J.~S. Marron (2008).
\newblock Statistical significance of clustering for high-dimension,
  low–sample size data.
\newblock {\em J. Am. Stat. Assoc.\/}~{\em 103\/}(483), 1281--1293.

\bibitem[\protect\citeauthoryear{Pan, Wang, Zhang, Zhu, and Zhu}{Pan
  et~al.}{2020}]{PWZZZ20}
Pan, W., X.~Wang, H.~Zhang, H.~Zhu, and J.~Zhu (2020).
\newblock Ball covariance: A generic measure of dependence in banach space.
\newblock {\em J. Am. Stat. Assoc.\/}~{\em 115\/}(529), 307--317.

\bibitem[\protect\citeauthoryear{Paul}{Paul}{2007}]{PA07}
Paul, D. (2007).
\newblock Asymptotics of sample eigenstructure for a large dimensional spiked
  covariance model.
\newblock {\em Stat. Sin.\/}~{\em 17\/}(4), 1617--1642.

\bibitem[\protect\citeauthoryear{Pavlenko, Björkström, and
  Tillander}{Pavlenko et~al.}{2012}]{PBT12}
Pavlenko, T., A.~Björkström, and A.~Tillander (2012).
\newblock Covariance structure approximation via glasso in high-dimensional
  supervised classification.
\newblock {\em J. Appl. Stat.\/}~{\em 39\/}(8), 1643--1666.

\bibitem[\protect\citeauthoryear{{R Core Team}}{{R Core
  Team}}{2022}]{RSoftware}
{R Core Team} (2022).
\newblock {\em R: A Language and Environment for Statistical Computing}.
\newblock Vienna, Austria: R Foundation for Statistical Computing.

\bibitem[\protect\citeauthoryear{Rothman, Levina, and Zhu}{Rothman
  et~al.}{2009}]{RLZ09}
Rothman, A.~J., E.~Levina, and J.~Zhu (2009).
\newblock Generalized thresholding of large covariance matrices.
\newblock {\em J. Am. Stat. Assoc.\/}~{\em 104\/}(485), 177--186.

\bibitem[\protect\citeauthoryear{Schott}{Schott}{2005}]{SC05}
Schott, J.~R. (2005).
\newblock Testing for complete independence in high dimensions.
\newblock {\em Biometrika\/}~{\em 92\/}(4), 951--956.

\bibitem[\protect\citeauthoryear{Schwarz}{Schwarz}{1978}]{SC78}
Schwarz, G. (1978).
\newblock Estimating the dimension of a model.
\newblock {\em Ann. Stat.\/}~{\em 6\/}(2), 461--464.

\bibitem[\protect\citeauthoryear{Shen, Priebe, and Vogelstein}{Shen
  et~al.}{2020}]{SPV20}
Shen, C., C.~E. Priebe, and J.~T. Vogelstein (2020).
\newblock From distance correlation to multiscale graph correlation.
\newblock {\em J. Am. Stat. Assoc.\/}~{\em 115\/}(529), 280--291.

\bibitem[\protect\citeauthoryear{Shen, Shen, and Marron}{Shen
  et~al.}{2013}]{SSM13}
Shen, D., H.~Shen, and J.~S. Marron (2013).
\newblock Consistency of sparse {PCA} in high dimension, low sample size
  contexts.
\newblock {\em J. Multivar. Anal.\/}~{\em 115}, 317--333.

\bibitem[\protect\citeauthoryear{Shen and Huang}{Shen and Huang}{2008}]{SH08}
Shen, H. and J.~Z. Huang (2008).
\newblock Sparse principal component analysis via regularized low rank matrix
  approximation.
\newblock {\em J. Multivar. Anal.\/}~{\em 99\/}(6), 1015--1034.

\bibitem[\protect\citeauthoryear{Srivastava and Reid}{Srivastava and
  Reid}{2012}]{SR12}
Srivastava, M.~S. and N.~Reid (2012).
\newblock Testing the structure of the covariance matrix with fewer
  observations than the dimension.
\newblock {\em J. Multivar. Anal.\/}~{\em 112}, 156--171.

\bibitem[\protect\citeauthoryear{Sz{\'e}kely and Rizzo}{Sz{\'e}kely and
  Rizzo}{2013}]{SR13}
Sz{\'e}kely, G.~J. and M.~L. Rizzo (2013).
\newblock The distance correlation t-test of independence in high dimension.
\newblock {\em J. Multivar. Anal.\/}~{\em 117}, 193--213.

\bibitem[\protect\citeauthoryear{Sz{\'e}kely, Rizzo, and Bakirov}{Sz{\'e}kely
  et~al.}{2007}]{SRB07}
Sz{\'e}kely, G.~J., M.~L. Rizzo, and N.~K. Bakirov (2007).
\newblock {Measuring and testing dependence by correlation of distances}.
\newblock {\em Ann. Stat.\/}~{\em 35\/}(6), 2769--2794.

\bibitem[\protect\citeauthoryear{Tan, Witten, and Shojaie}{Tan
  et~al.}{2015}]{TWS15}
Tan, K.~M., D.~Witten, and A.~Shojaie (2015).
\newblock The cluster graphical lasso for improved estimation of gaussian
  graphical models.
\newblock {\em Comput. Stat. Data Anal.\/}~{\em 85}, 23--36.

\bibitem[\protect\citeauthoryear{Tibshirani}{Tibshirani}{1996}]{TI96}
Tibshirani, R. (1996).
\newblock Regression shrinkage and selection via the lasso.
\newblock {\em J. R. Stat. Soc. B\/}~{\em 58\/}(1), 267--288.

\bibitem[\protect\citeauthoryear{Wang, Li, and Leng}{Wang et~al.}{2009}]{WLL09}
Wang, H., B.~Li, and C.~Leng (2009).
\newblock Shrinkage tuning parameter selection with a diverging number of
  parameters.
\newblock {\em J. R. Stat. Soc. B\/}~{\em 71\/}(3), 671--683.

\bibitem[\protect\citeauthoryear{Wang, Li, and Tsai}{Wang et~al.}{2007}]{WLT07}
Wang, H., R.~Li, and C.-L. Tsai (2007, 08).
\newblock Tuning parameter selectors for the smoothly clipped absolute
  deviation method.
\newblock {\em Biometrika\/}~{\em 94\/}(3), 553--568.

\bibitem[\protect\citeauthoryear{Wang, Kim, and Li}{Wang et~al.}{2013}]{WKL13}
Wang, L., Y.~Kim, and R.~Li (2013).
\newblock Calibrating nonconvex penalized regression in ultra-high dimension.
\newblock {\em Ann. Stat.\/}~{\em 41\/}(5), 2505--2536.

\bibitem[\protect\citeauthoryear{Witten, Tibshirani, and Hastie}{Witten
  et~al.}{2009}]{WTH09}
Witten, D.~M., R.~Tibshirani, and T.~A. Hastie (2009).
\newblock A penalized matrix decomposition, with applications to sparse
  principal components and canonical correlation analysis.
\newblock {\em Biostatistics\/}~{\em 10\/}(3), 515--534.

\bibitem[\protect\citeauthoryear{Yamada, Hyodo, and Nishiyama}{Yamada
  et~al.}{2017}]{YHN17}
Yamada, Y., M.~Hyodo, and T.~Nishiyama (2017).
\newblock Testing block-diagonal covariance structure for high-dimensional data
  under non-normality.
\newblock {\em J. Multivar. Anal.\/}~{\em 155}, 305--316.

\bibitem[\protect\citeauthoryear{Yang, Ma, and Buja}{Yang et~al.}{2014}]{YMB14}
Yang, D., Z.~Ma, and A.~Buja (2014).
\newblock A sparse singular value decomposition method for high-dimensional
  data.
\newblock {\em J. Comput. Graph. Stat.\/}~{\em 23\/}(4), 923--942.

\bibitem[\protect\citeauthoryear{Yata and Aoshima}{Yata and
  Aoshima}{2016}]{YA16}
Yata, K. and M.~Aoshima (2016).
\newblock High-dimensional inference on covariance structures via the extended
  cross-data-matrix methodology.
\newblock {\em J. Multivar. Anal.\/}~{\em 151}, 151--166.

\bibitem[\protect\citeauthoryear{Yu, Wang, and Samworth}{Yu
  et~al.}{2015}]{YWS15}
Yu, Y., T.~Wang, and R.~J. Samworth (2015).
\newblock {A useful variant of the Davis—Kahan theorem for statisticians}.
\newblock {\em Biometrika\/}~{\em 102\/}(2), 315--323.

\bibitem[\protect\citeauthoryear{Zhu, Zhang, Yao, and Shao}{Zhu
  et~al.}{2020}]{ZZYS20}
Zhu, C., X.~Zhang, S.~Yao, and X.~Shao (2020).
\newblock {Distance-based and RKHS-based dependence metrics in high dimension}.
\newblock {\em Ann. Stat.\/}~{\em 48\/}(6), 3366--3394.

\bibitem[\protect\citeauthoryear{Zou, Hastie, and Tibshirani}{Zou
  et~al.}{2006}]{ZHT06}
Zou, H., T.~Hastie, and R.~Tibshirani (2006).
\newblock Sparse principal component analysis.
\newblock {\em J. Comput. Graph. Stat.\/}~{\em 15\/}(2), 265--286.

\bibitem[\protect\citeauthoryear{Zou, Hastie, and Tibshirani}{Zou
  et~al.}{2007}]{ZHT07}
Zou, H., T.~Hastie, and R.~Tibshirani (2007).
\newblock {On the “degrees of freedom” of the lasso}.
\newblock {\em Ann. Stat.\/}~{\em 35\/}(5), 2173--2192.

\end{thebibliography}
\end{document}